\documentclass{JAC2003}

\usepackage{graphicx}
\usepackage{booktabs}

\begin{document}
\title{TARGET STUDIES FOR THE PRODUCTION OF LITHIUM8 FOR NEUTRINO PHYSICS USING A LOW ENERGY CYCLOTRON}

\author{Adriana Bungau, Roger Barlow, University of Huddersfield, Huddersfield, UK\\
Michael Shaevitz, Columbia University, New York, USA\\
Janet Conrad, Joshua Spitz, Massachusetts Institute of Technology, Massachusetts, USA}

\maketitle

\begin{abstract}
  Lithium 8 is a short lived beta emitter producing a high energy anti-neutrino, which is very suitable for making several measurements of fundamental quantities. It is proposed to produce Lithium 8 with a commercially available 60 MeV cyclotron using protons or alpha particles on a Beryllium 9 target. We have used the GEANT4 program to model these processes, and calculate the anti-neutrino fluxes that could be obtained in a practical system. We also calculate the production of undesirable contaminants such as Boron 8, and show that these can be reduced to a very low level.
\end{abstract}

\section{INTRODUCTION}

IsoDAR (isotropic decay at-rest) sources of electron-flavor neutrino beams have been considered in the past  for underground physics~\cite{Davison}~\cite{Lutostansky}, however the problem has been to generate isotopes at sufficiently high rates to meet the required physics goals for beyond Standard Model searches. Modern improvements in high power cyclotrons have opened up the opportunity for construction of a sufficiently powerful isoDAR source for the first time. 

Two cyclotron designs originated as injectors for the DAE$\delta$ALUS (Decay-At-rest Experiment
for $\delta_{CP}$ studies At a Laboratory for Underground Science), CP-violation experiment were proposed. The base design uses a traditional compact cyclotron to direct protons from a high intensity H$_2^+$ source, with 60 MeV/n, onto a low-A target such as beryllium. The H$_2^+$ is chosen to reduce space charge and minimise beam losses at extraction. The accelerator must achieve a peak current of 5 mA. The second design proposes an RFQ injection of  H$_2^+$ into a small separated cyclotron.
An ion source is required to feed the isoDAR cyclotron and a prototype ion source that supplies
sufficient beam has already been constructed. Closely associated with the cyclotron design for isoDAR is the target/shielding development. The target is designed to maximise the anti-neutrinos from isotopes that decay-at-rest.  The isotopes are produced by the following interaction:

\begin{equation}\label{eq:isotopes}
    p + ^{9}Be \rightarrow ^{8}Li + 2p
\end{equation}

 This will produce isotopes that subsequently $\beta^{-}$-decay with high $Q$-value. The KamLAND detector has been chosen for the isoDAR source. The detector has a radius of 13 m and it's centre is situated at 16 m from the face of the target. The isoDAR parameters are summarised in Table~\ref{l2ea4-t1}.

\begin{table}[hbt]
   \centering
   \caption{IsoDAR Parameters}
   \begin{tabular}{lcc}
       \toprule
       \textbf{Parameter} & \textbf{Value}  \\ 
       \midrule
           Accelerator      & 60 MeV/amu of    H$_2^+$          \\
           Current     & 10 mA of p on target                 \\
           Power        & 600 kW                 \\
           Protons/year       & 1.97$\times10^{24}$                  \\
           Duty cycle & 90$\%$ \\
           Run period & 5 years\\
           $\overline{\nu}_{e}$-electron event total& 7256\\
           Visible energy threshold & 3 MeV\\
           Expected $sin^{2}\theta_{W}$ 1$\sigma$ precision & 0.00376\\
       \bottomrule
   \end{tabular}
   \label{l2ea4-t1}
\end{table}

\section{COMPUTATION DETAILS}
In the present work, the GEANT4 simulation code~\cite{GEANT4} has been used to simulate the low-energy protons induced isotopes production in Be targets. GEANT4 provides an extensive set of hadronic physics models for energies up to 10 - 15 GeV, both for the intra-nuclear cascade region and for modelling of evaporation. In other codes, like MCNP for example, the Bertini model is used by default for nucleons and pions, while the ISABEL model is used for other particle types~\cite{mcnpx}. The Bertini model does not take into account the nuclear structure effects in the inelastic interactions during the intranuclear cascade and therefore the code modelling of interactions at energies much below 100 MeV is questionable~\cite{spallation_paper}. On the other hand, GEANT4 does not offer default models, it is the user's responsibility to select the appropriate model for each specific application. One of the theory driven intra-nuclear cascade models provided by GEANT4 is the Binary Cascade model~\cite{Binary}. In this model the propagation through the nucleus of the incident hadron and the secondaries it produces is modelled by a cascade series of two-particle collision, hence the name binary cascade. Between collisions the hadrons are transported in the field of the nucleus by Runge-Kutta method. The model is valid for incident protons, neutrons and pions and it reproduces detailed proton and neutron cross section data in the region 0-10 GeV. For incident particle energies below 150 MeV, the Binary Cascade will quickly call the G4Precompund model, which will handle the proton induced isotopes production inside the Be target. 

Also, in GEANT4 there are no tracking cuts. All particles are tracked up to zero range. In order to accurately simulate the interactions of the low-energy spallation neutrons inside the target, the QGSP\_BIC\_HP physics list has been used. In this study we also incorporated the effects of the S($\alpha,\beta$) coefficient which takes into account the atomic translational motion, vibration and rotation effects for thermal neutrons scattering. All the neutron cross-section data libraries are included as well.

\section{TARGET SIMULATIONS}
\begin{figure}[h]
    \centering
    \includegraphics*[width=80mm]{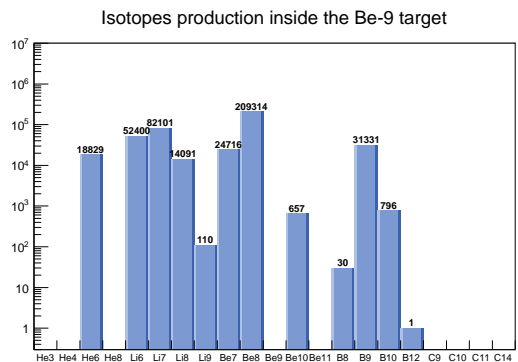}
    \caption{GEANT4 simulation of isotopes production inside the Be target.}
    \label{simple-target-iso}
\end{figure}

In this study beryllium has been chosen as the target material. Compared with other light target materials, beryllium has a higher melting point and thermal conductivity, and is less reactive with air. Also, beryllium has the smallest binding energy for neutrons of any stable element, at 1.67 MeV. So when the incident proton interacts, it will knock out a neutron of roughly the same energy. A cylindrical $^{9}$Be target  having a diameter of 20 cm and a length of 20 cm has been implemented into the simulation code. Each simulation consists of $10^{7}$ incident protons with a kinetic energy of 60 MeV. The simulation output consists of the total number of isotopes produced inside the target, and it can been seen in Fig.~\ref{simple-target-iso} that the ratio of $^{8}$Li to $^{8}$B is approximately 470:1.

\begin{figure}[h!]
    \centering
    \includegraphics*[width=60mm, height=55mm]{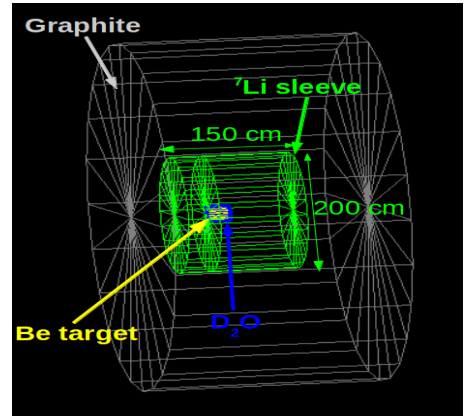}
    \caption{GEANT4 simulation of a cylindrical Be target surrounded by D$_{2}$O moderator, Li sleeve and graphite reflector.}
    \label{geometry}
\end{figure}

From the total number of $^{8}$Li isotopes produced in the target, 63.8\% are produced in proton inelastic processes and 35.9\% are produced in neutron inelastic interactions. The rest 0.3\% $^{8}$Li isotopes are due to other processes.

A large fraction of the secondary neutrons exit the target and are lost. In order to make a better use of these neutrons, a Li sleeve has been modelled surrounding the Be target, with a 5 cm layer of heavy water in between, in order to moderate the neutrons down to the energy we need for the neutron capture to take place inside the sleeve. The Li sleeve is surrounded by a graphite neutron reflector, in order to reflect back into the sleeve the outgoing neutrons. The Be target dimensions were kept the same as before. The reflector radius is 3 m and it is 3 m long. The design is shown in Fig.~\ref{geometry}.

\begin{figure}[h]
    \centering
    \includegraphics*[width=75mm, height=60mm]{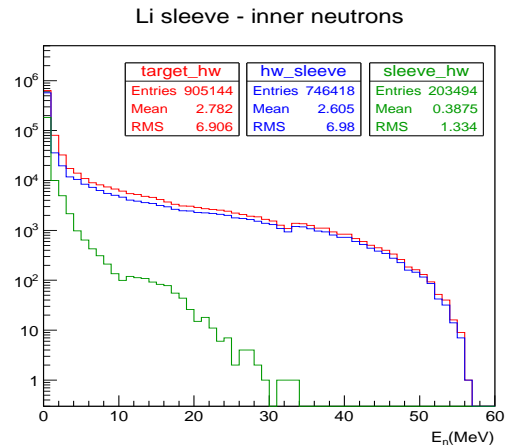}
    \caption{Energy spectra of the neutrons inside the sleeve.}
    \label{inner_neutrons}
\end{figure}

\begin{figure}[h]
    \centering
    \includegraphics*[width=75mm, height=60mm]{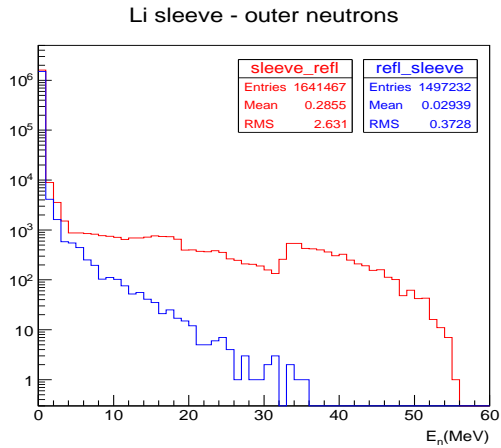}
    \caption{Energy spectra of the neutrons exiting the sleeve.}
    \label{outer_neutrons}
\end{figure}

The energy spectra have been recorded in the simulation for all the neutrons as they pass from one volume to another. The inner neutrons are the neutrons inside the Li sleeve, and the outer neutrons, those outside the sleeve. Figure~\ref{inner_neutrons} shows the energy of the neutrons as they exit the target and enter the D$_{2}$O moderator, and also for the neutrons leaving the moderator volume and entering the sleeve, and those reflected back. The spectra of the neutrons as they exit the sleeve and enter the graphite reflector, and those reflected back, are shown in Fig.~\ref{outer_neutrons}.

\begin{figure}[h]
    \centering
    \includegraphics*[width=80mm]{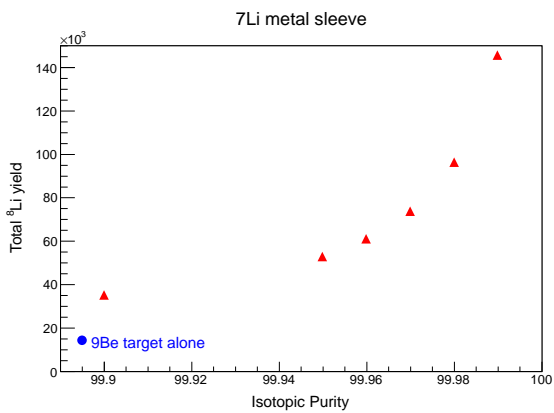}
    \caption{The total $^{8}$Li yield inside the Be target and the surrounding metallic Li sleeve for different isotopic purities inside the sleeve.}
    \label{purities}
\end{figure}

Initially, the Li sleeve was designed to be 110 cm long starting from the front of the target. The Li sleeve has been extended 40 cm in front of the Be target, in order to capture the spallation neutrons emitted backwards with respect to the incident proton beam. The metallic Li sleeve consists of the two stable Li isotopes: $^{6}$Li and $^{7}$Li and we found that the $^{8}$Li yield has a strong dependence upon the purity of $^{7}$Li inside the sleeve. The benefits of having the surrounding metallic Li sleeve can be observed for a $^{7}$Li purity above 99.9\%, as shown in Fig.~\ref{purities}.

\begin{figure}[h]
    \centering
    \includegraphics*[width=80mm, height=60mm]{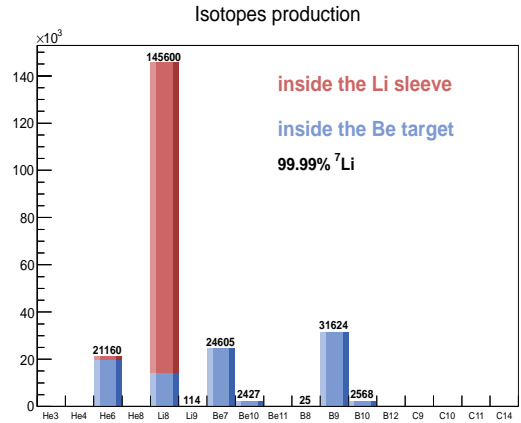}
    \caption{GEANT4 simulation of the isotopes production inside the Be target and the surrounding $^{7}$Li metal sleeve.}
    \label{total-yields}
\end{figure}

The stacked histogram of the isotopes yield both inside the Be target
and the surrounding sleeve is shown in Fig.~\ref{total-yields} for a
$^{7}$Li purity of 99.99\%. The fraction of the $^{8}$Li yield which is produced inside the Be target is 9.7\% of the total $^{8}$Li yield. It should be noted here that there is a big difference in the prediction mechanisms for the $^{8}$Li production rates inside the target and the sleeve. The rate in the target is predicted using theoretical models which have been validated in the past using neutron yield data in various targets, including Be. While these models predict reasonably well the neutron yield, they have not been validated for proton induced isotopes production. Once the neutrons are produced inside the target, they are simulated using the data-driven high precision (HP) models. On the other hand, the rate in the sleeve is predicted using these data-driven models, which include and use all the available neutron xsection libraries.

\section{CONCLUSION}
In this paper we investigated different geometries in order to find the optimum setup with respect to maximising the production of $^{8}$Li isotopes. Starting with a simple beryllium target, we have increased the total yield by a factor of 10, using a 99.99\% pure $^{7}$Li metallic sleeve surrounding the target in order to capture the outgoing spallation neutrons. A heavy water moderator has been added around the target to moderate the neutrons down to the energy needed for neutron capture inside the sleeve and a graphite reflector surrounding the sleeve was placed to reflect the neutrons back into the sleeve. With the current geometry, the total $^{8}$Li yield is 145600 isotopes for $10^{7}$ protons on target.

\end{document}